\documentclass[11pt]{article}
\usepackage{moriond,epsfig}
\usepackage{amsmath, amssymb}

\def\Journal#1#2#3#4{{#1} {\bf #2}, #3 (#4)}

\def\NPB{{\em Nucl. Phys.} B}
\def\PLB{{\em Phys. Lett.}  B}
\def\PRL{\em Phys. Rev. Lett.}
\def\PRD{{\em Phys. Rev.} D}

\def\be{\begin{equation}}
\def\ee{\end{equation}}
\def\bea{\begin{eqnarray}}
\def\eea{\end{eqnarray}}

\begin{document}
\vspace*{4cm}
\title{LFV radiative Decays and Leptogenesis in the SUSY seesaw model
\footnote{Presentation given by T.~Shindou}}

\author{ S.~T.~Petcov \footnote{Also at: Institute of
Nuclear Research and Nuclear Energy,
Bulgarian Academy of Sciences, 1784 Sofia, Bulgaria.} and 
T.~Shindou}%\footnote{Presented by T.~Shindou.}}

\address{
 Scuola Internazionale Superiore di Studi Avanzati, and \\
% I-34014 Trieste, Italy\\ }
Istituto Nazionale di Fisica Nucleare, I-34014 Trieste, Italy }

\maketitle\abstracts{
The lepton flavour violating charged lepton decays $\mu \to e + \gamma$
and thermal leptogenesis are analysed 
in the minimal supersymmetric standard model with see-saw mechanism 
of neutrino mass generation and soft supersymmetry
breaking terms with universal boundary conditions.
Hierarchical spectrum of heavy Majorana neutrino masses, 
$M_1 \ll  M_2 \ll M_3$, is considered.  
In this scenario, the requirement of successful thermal leptogenesis implies 
a lower bound on $M_1$.
For the natural GUT values of the heaviest right-handed 
Majorana neutrino mass,
$M_3 \gtrsim 5\times 10^{13}$ GeV, and supersymmetry particle masses
in the few $\times$ 100 GeV range,
the predicted $\mu\to e+\gamma $ decay rate exceeds by few order
of magnitude the experimental upper limit.
This problem is avoided if the matrix of neutrino Yukawa couplings has
a specific structure.
The latter leads to a correlation between the 
baryon asymmetry of the Universe predicted by leptogenesis, 
$\text{BR}(\mu\to e+\gamma)$
and the effective Majorana mass in neutrinoless double beta
decay.
}

\vspace{-5mm}
\section{Introduction}
\vspace{-3mm}

The supersymmetric (SUSY) extension of the Standard Model (SM) is 
a widely discussed candidate of a theory beyond the SM.
If the SUSY particles have masses in the few $\times$ 100 GeV range, 
they will be observed in the large hadron collider (LHC) experiments.
In this case lepton flavour violating (LFV) processes, like 
$\mu\to e+\gamma$ decay, etc., are also predicted to take place 
with rates which can be close to the existing experimental upper limits.

We focus on the minimal supersymmetric standard
model with right-handed heavy Majorana neutrinos (MSSMRN).
In this model one can implement the seesaw mechanism \cite{seesaw} 
which provides a natural explanation of the smallness 
of neutrino masses.
The seesaw mechanism predicts the light massive 
neutrinos to be Majorana particles. In this case
the process of neutrinoless double beta 
($0\nu 2\beta$) decay, $(A,Z)\to (A,Z+2)+e^-+e^-$, can occur. 
The seesaw mechanism provides also,
through the leptogenesis scenario \cite{LeptoG}, an attractive 
explanation of the observed baryon asymmetry of the Universe (BAU).

In MSSMRN \footnote{In the SM with massive neutrinos, 
$\text{BR}(\mu\to e+\gamma)$ is suppressed by the factor \cite{SP76,BPP77}
$(m_{j}/M_W)^4 < 6.7\times 10^{-43}$.},
the neutrino Yukawa couplings which do not conserve the 
lepton flavour,
affect the predictions of LFV processes \cite{BorMas86}.
Even when the soft SUSY breaking terms present at the cutoff scale ($M_X$) 
are flavour blind,
non-zero flavour mixing in the slepton sector is 
induced through the renormalization group (GR) running 
between $M_X$ and the seesaw scale $M_R<M_X$.
For $M_R\gtrsim 5\times 10^{13}$ GeV one typically gets in SUSY GUT \cite{gut},
the prediction for the $\mu\to e+\gamma$
decay branching ratio, $\text{BR}(\mu\to e+\gamma)$, in MSSMRN 
with SUSY particles masses in the few $\times$ 100 GeV range, 
is in conflict with the present experimental upper bound \cite{mega},
$\text{BR}(\mu\to e+\gamma)< 1.2\times 10^{-11}$.
Thus, the leading contribution(s) to $\text{BR}(\mu\to e+\gamma)$ has 
to be suppressed.
%This upper limit will be improved in the future experiment.
% In the experiment MEG under preparation at PSI\cite{psi} 
%it is planned to reach a sensitivity to
%\begin{equation}
%\text{B}(\mu \ra e+\gamma)\sim (10^{-13} - 10^{-14})\,.
%\end{equation}
%Then the measurement of LFV gives a very big hint on the structure 
%of the neutrino Yukawa coupling matrix.

In this article, we consider a specific form of the matrix of neutrino Yukawa
couplings, $\mathbf{Y_N}$, for which
$\text{BR}(\mu\to e+\gamma)$ is suppressed, but still can be within 
the sensitivity of the ongoing MEG experiment.
The form thus chosen of $\mathbf{Y_N}$ implies 
a correlation between the predicted values of 
the effective Majorana mass in $0\nu 2\beta$ decay,
of BAU, and
of $\text{BR}(\mu\to e+\gamma)$.

\vspace{-2mm}
\section{Seesaw model and a parametrisation neutrino Yukawa coupling matrix}
\vspace{-3mm}

We consider MSSMRN. The presence of the RH neutrinos, $N^c_i$, in the
theory makes it possible to introduce 
neutrino Yukawa couplings and Majorana mass term for $N^c_i$
in the superpotential.
In the framework of MSSMRN, a basis in which 
both the charged lepton mass matrix, $\mathbf{Y_E}$, and the 
heavy Majorana neutrino mass matrix, $\mathbf{M_N}$, are real and diagonal,
can always be chosen without loss of generality. Henceforth,
we will work in this basis and denote the diagonal RH neutrino mass
matrix by $\mathbf{D_N}\equiv \mathrm{diag}(M_1,M_2,M_3)$.

Below the seesaw scale, $M_R = {\rm min}(M_j)$, the heavy RH neutrino
fields $N_j$ are integrated out, and as a result of the electroweak
symmetry breaking, Majorana mass term for 
the left-handed flavour neutrinos is generated. The corresponding
mass matrix can be expressed in terms of the matrix of neutrino Yukawa couplings,
$\mathbf{Y_N}$, and $\mathbf{M_N}$ as
%%%%%%%%%%%%%%%%%%%%%%%%%%%
\begin{align}
(m_{\nu})^{ij} =(\mathbf{U}^*)^{ik}m_k(\mathbf{U}^{\dagger})^{kj}
=v_u^2~(\mathbf{Y_{N}^T})^{ik}(\mathbf{M_N^{-1}})^{kl}(\mathbf{Y_N})^{lj}\;.
\label{Match_at_MR}
\end{align}
%%%%%%%%%%%%%%%%%%%%%%%%%%%%
%
Here $v_u = v \sin\beta$, where $v = 174$ GeV, $\tan\beta$ is the
ratio of the vacuum expectation values of up-type and down-type Higgs
fields, and $\mathbf{U}$ is the Pontecorvo--Maki--Nakagawa--Sakata (PMNS) mixing
matrix \cite{BPont57}, 
\begin{equation}
\mathbf{U}=
 \begin{pmatrix}
 c_{12} c_{13} & s_{12} c_{13} & s_{13} e^{-i \delta}\\
 -s_{12} c_{23} - c_{12} s_{23} s_{13} e^{i \delta}
 & c_{12} c_{23} - s_{12} s_{23} s_{13} e^{i \delta} & s_{23} c_{13}\\
 s_{12} s_{23} - c_{12} c_{23} s_{13} e^{i \delta} &
 - c_{12} s_{23} - s_{12} c_{23} s_{13} e^{i \delta} & c_{23} c_{13}
\end{pmatrix}
{\rm diag}(1, e^{i \frac{\alpha}{2}}, e^{i \frac{\beta_M}{2}}) \, .
\end{equation}
Eq.~(\ref{Match_at_MR}) can be ``solved'' as \cite{Iba01}
\begin{equation}
\mathbf{Y_N}=\frac{1}{v_u}\sqrt{\mathbf{D_N}}\mathbf{R}
\sqrt{\mathbf{D}_{\nu}}\mathbf{U}^{\dagger}\;,
\label{sol_YN}
\end{equation}
where $\mathbf{D_{\nu}}=\mathrm{diag}(m_1,m_2,m_3)$ and 
$\mathbf{R}$ is a complex orthogonal matrix, 
$\mathbf{R}^T\mathbf{R}=\mathbf{1}$.
The CP phases in $\mathbf{R}$ are directly related to the CP asymmetry 
parameter in leptogenesis. In addition, these phases 
can affect significantly the predicted rates of the LFV 
processes \cite{pst,prst,dprr}.

\vspace{-2mm}
\section{Models with texture zero for the inverted hierarchical light neutrinos}
\vspace{-3mm}
Hereafter we focus on the case of light neutrino mass spectrum of
inverted hierarchical (IH) \\ type \footnote{
The cases of normal hierarchical and of quasi-degenerate light neutrinos are
discussed in Ref.~14.
},
$m_1\sim m_2\simeq \sqrt{|\Delta m_{31}^2|}\simeq 0.05$ eV, and $m_3\simeq 0$.
For IH light neutrinos, the effective Majorana mass
in $0\nu 2\beta$ decay, $\langle m\rangle$, is given by 
\begin{equation}
\left|\langle m\rangle\right| = \left|
m_1c_{12}^2c_{13}^2+m_2s_{12}^2c_{13}^2e^{i\alpha}\right|\;,
\label{0nu2beta}
\end{equation}
$\theta_{12}$ and $\theta_{13}$ being the solar and CHOOZ mixing angles,
respectively.
Thus, $\sqrt{|\Delta m_{31}^2|}\cos 2\theta_{12}\lesssim |\langle m\rangle|
\lesssim 
\sqrt{|\Delta m_{31}^2|}$, the two limits corresponding to the 
CP-conserving values of $\alpha=0$;$\pi$.

\begin{figure}
%\rule{5cm}{0.2mm}\hfill\rule{5cm}{0.2mm}
%\vskip 2.5cm
%\rule{5cm}{0.2mm}\hfill\rule{5cm}{0.2mm}
\begin{center}
\psfig{figure=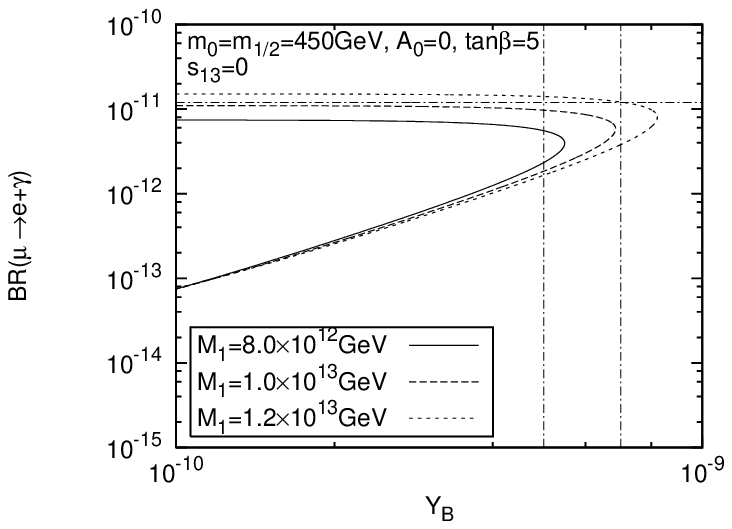,height=1.65in, width=3.0in}
\psfig{figure=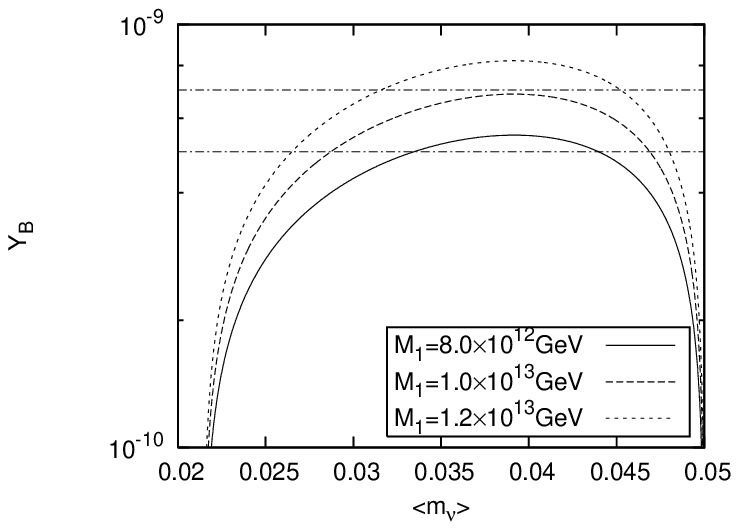,height=1.65in, width=3.0in}
\end{center}
\caption{The correlation between $Y_B$, 
$\text{B}(\mu\to e+\gamma)$ and $\langle m\rangle$ for the case A
($(\mathbf{Y_N})_{21}=0$).}
\label{fig:rel3values}
\end{figure}

Even when we consider MSSMRN with flavour universal soft scalar masses, $m_0$,
GUT gaugino mass, $m_{1/2}$, and universal $A$-term coefficient, $A_0$,
at the cutoff scale $M_X$,
slepton flavour mixing leading to 
LFV decays such as $\mu\to e+\gamma$, 
is induced by
$\mathbf{Y_N}$ which does not conserve lepton flavour, 
through the RG running between $M_R$ and $M_X$.
The branching ratio $\text{BR}(\mu\to e+\gamma)$ in this framework is predicted as
\begin{equation}
\text{BR}(\mu\to e+\gamma)\simeq
\frac{\alpha^3}{G_F^2}
\frac{\left|(\mathbf{\tilde{M}_L^2})_{21}\right|^2}{m_S^8}
\frac{\langle h_d\rangle^2}{\langle h_u\rangle^2}
\propto |(\mathbf{Y_N^{\dagger}LY_N})_{21}|^2\;.
\end{equation}
Here
$\mathbf{L}=\mathrm{diag}(\ln\frac{M_1}{M_X},\ln\frac{M_2}{M_X},\ln\frac{M_3}{M_X})$ and
$m_S$ denotes the SUSY particle mass scale which 
can be approximately estimated \cite{ppty} as
\begin{equation}
m_S^8\simeq 0.5 m_0^2 M_{1/2}^2(m_0^2+0.6 M_{1/2}^2)^2\;.
\end{equation}
Using Eq.~(\ref{sol_YN}), one can decompose
$(\mathbf{Y_N^{\dagger}LY_N})_{21}$ as
\begin{align}
(Y_N^{\dagger}LY_N)_{21}=&
\frac{M_3}{v_u^2}\ln\frac{M_3}{M_X}
(\mathbf{R}\sqrt{D_{\nu}}U^{\dagger})^*_{32}
(\mathbf{R}\sqrt{D_{\nu}}U^{\dagger})_{31}
+\frac{M_2}{v_u^2}\ln\frac{M_2}{M_X}
(\mathbf{R}\sqrt{D_{\nu}}U^{\dagger})^*_{22}
(\mathbf{R}\sqrt{D_{\nu}}U^{\dagger})_{21}\nonumber\\
&+\frac{M_1}{v_u^2}\ln\frac{M_1}{M_X}
(\mathbf{R}\sqrt{D_{\nu}}U^{\dagger})^*_{12}
(\mathbf{R}\sqrt{D_{\nu}}U^{\dagger})_{11}\;.
\end{align}
If RH neutrinos have hierarchical mass spectrum, $M_1\ll M_2\ll M_3$,
the term proportional to $M_3$ gives the dominant contribution \cite{prst,dprr}.
Suppressing this term is the most natural way
to avoid too large SUSY contribution to $\text{BR}(\mu\to e+\gamma)$.
This is realised by taking matrix $\mathbf{R}$ in the form \cite{prst},
\begin{equation}
\mathbf{R}=\begin{pmatrix}
	\cos\omega&\sin\omega&0\\
	-\sin\omega&\cos\omega&0\\
	0&0&1
\end{pmatrix}\;,
\end{equation}
where $\omega$ is a complex angle -- the leptogenesis CPV parameter. However, in this case, 
the requirement of  successful leptogenesis 
leads to a very strong constraint on $M_1$ :
$M_1\geq 6.7\times 10^{12}$. 
The latter implies $M_2\gtrsim 3\times 10^{13}$ GeV,
which in turn makes the contribution of the term 
$\propto M_2$ in $\text{BR}(\mu\to e+\gamma)$ bigger than the experimental 
upper bound by $\sim$ 3 orders of magnitude in the case the SUSY
particles have masses in the few $\times$ 100 GeV range.
Thus, we consider a scheme in which the $M_2$ contribution to
$\text{BR}(\mu\to e+\gamma)$
is also suppressed \cite{text0}.
There are two possibilities:
\begin{itemize}
\item[A:] $(\mathbf{Y_N})_{21}=0$, which is satisfied for
$\tan\omega=\tan\theta_{12}e^{-i\alpha/2}$.
\item[B:] $(\mathbf{Y_N})_{22}=0$, which holds if
$\tan\omega=-\cot\theta_{12}e^{-i\alpha/2}+\mathcal{O}(s_{13})$.
\end{itemize}
Since in the cases A and B
the complex parameter $\omega$ is completely determined
by the low energy neutrino mixing parameters, $\theta_{12}$ and 
the Majorana phase $\alpha$,
the quantities $Y_B$, $\text{BR}(\mu\to e+\gamma)$,
and $\langle m\rangle$
are related to each other through the Majorana CP phase $\alpha$.
The correlation between these three observables for Case A is 
illustrated in Fig.~\ref{fig:rel3values} (for further details, see Ref.~14.).
%In the both cases, the complex parameter $\omega$ can be completely determined
%by the low energy neutrino parameters. 
%For both case, the baryon asymmetry of the universe $Y_B$ and 
%$|(\mathbf{Y_N^{\dagger}Y_N})_{21}|^2$ which corresponds to the
%$\text{BR}(\mu\to e+\gamma)$ are estimated as
%\begin{equation}
%Y_B\simeq -6.6\times 10^{-14}\left(\frac{M_1}{10^{9}\text{GeV}}\right)
%\left(\frac{0.05\text{eV}}{m_2}\right)^{0.16}\sin\alpha\;,
%\end{equation}
%and 
%\begin{equation}
%\left|(\mathbf{Y_N^{\dagger}Y_N})_{21}\right|^2\simeq 
%\left(\frac{M_1L_1\sqrt{|\Delta m_{31}^2|}}{v_u^2}\right)^2
%4\sin^2\frac{\alpha}{2}c_{23}^2c_{12}^2s_{12}^2\;,
%\end{equation}
%respectively. Therefore it is easily seen that 
%$Y_B$, $\text{BR}(\mu\to e+\gamma)$,
%and $\langle m\rangle$ in Eq.~(\ref{0nu2beta}) 
%correspond to each other through the Majorana CP phase $\alpha$.

\vspace{-2mm}
\section{Summary}
\vspace{-3mm}

The experiments at LHC are planed to start in 2007 and
are the only experiments which could provide direct evidence for 
new physics beyond the SM.
If low energy scale SUSY is realised in nature,
supersymmetry (SUSY) particles might be
observed in the few $\times$ 100 GeV mass range.
In the SUSY extension of the SM incorporating the seesaw mechanism of neutrino
mass generation and SUSY particles masses in the few $\times$ 100 GeV range, 
the existing experimental limit on $\text{BR}(\mu\to e+\gamma)$ and the 
requirement of successful leptogenesis imply rather stringent constraint 
on the matrix of neutrino Yukawa couplings, $\mathbf{Y_N}$.
For hierarchical heavy Majorana neutrino
spectrum and inverted hierarchical light neutrino masses,
these constraints lead to a specific rather simple 
form of $\mathbf{Y_N}$.
The specific form of $\mathbf{Y_N}$ thus found implies
that the leptogenesis CPV parameter is given by the Majorana phase in 
the PMNS matrix $\mathbf{U}$.
Thus, the predicted values of the baryon asymmetry of the Universe,
of $\text{BR}(\mu\to e+\gamma)$, and of
the effective Majorana mass in the neutrinoless double beta decay, are all
correlated.

\vspace{-2mm}
\section*{Acknowledgements}
\vspace{-3mm}
This work was supported in part by the Italian
MIUR and INFN  under the programs
``Fisica Astroparticellare''.

\end{document}